\documentclass[12pt,letterpaper]{article}
\usepackage{jheppub}

\usepackage{color}

\usepackage{epsfig,multirow,subfigure} 
\usepackage{amscd}
\usepackage[matrix,arrow,curve]{xy}
\usepackage{verbatim}
\usepackage{latexsym}
\usepackage{amsfonts,amsthm,amsmath,mathtools}

\newcommand{\be}{\begin{equation}}
\newcommand{\ee}{\end{equation}}
\newcommand{\bea}{\begin{eqnarray}}
\newcommand{\eea}{\end{eqnarray}}
 
\subheader{\begin{flushright}
UCSD-PTH-12-16\\ 
IPPP/12/84\\
DCPT/12/168
\end{flushright}}
\title{\centering
BPS states and their reductions \\
}

\author[a]{Prarit Agarwal,}
\author[a,b]{Antonio Amariti,}
\author[c]{Alberto Mariotti,}
\author[d]{Massimo Siani.}

\affiliation[a]{Department of Physics, University of California, \\
San Diego La Jolla, CA 92093-0354, USA}
\affiliation[b]{Laboratoire de Physique Th\'eorique de l'\'Ecole Normale Sup\'erieure \\
and CNRS UMR 8549,
24 Rue Lhomond, Paris 75005, France}
\affiliation[c]{Institute for Particle Physics Phenomenology,\\
Department of Physics, Durham University, DH1 3LE, United Kingdom}
\affiliation[d]{Department of Particle Physics and Astrophysics \\
Weizmann Institute of Science, Rehovot 76100, Israel}

\emailAdd{pagarwal@physics.ucsd.edu}
\emailAdd{amariti@physics.ucsd.edu}
\emailAdd{alberto.mariotti@durham.ac.uk}
\emailAdd{massimo@weizmann.ac.il}

\abstract{
We develop a
method to identify the BPS states 
in the Hilbert space of a supersymmetric field theory
on a generic curved space
which preserves at least two real supercharges.
We also  propose a one-to-one map between
BPS states in $d$-dimensional field theories and states
that contribute to the supersymmetric partition function of a corresponding $(d-1)$-dimensional
field theory. 
As an application we 
obtain the 
superconformal index on rounded and 
squashed three spheres,
and we show a natural reduction of the respective indices
to the three-dimensional exact partition functions. 
We discuss the validity of the
correspondence  both at the perturbative and at the non-perturbative
level and exploit the idea to uplift the computation of the exact supersymmetric partition
function on a general manifold to a higher dimensional index.
}

\begin{document}

\maketitle

\section{Introduction}

Supersymmetric field theories on curved backgrounds are of great interest due to the fact that
they capture the full quantum information about quantities of the corresponding field theory defined
on flat space, where the same exact quantum results would be difficult to find.

Different choices of the background manifold correspond to a different information about the flat space theory.
One of the first examples has been $T^3 \times \mathbb{R}$ \cite{Witten:1982df}, where the supersymmetric
partition function counts supersymmetric vacua and has been dubbed index 
(see also \cite{Cecotti:1981fu}).
 Because it is  an integer number, it cannot depend upon the continuous superpotential and gauge couplings, under mild
assumptions.
More recently another manifold, the Euclidean $S^3 \times S^1$, has attracted much attention,
because in this case the supersymmetric partition function is an index that
counts a reduced set of states of the flat space theory, namely the BPS states \cite{Romelsberger:2005eg,Kinney:2005ej}.
The latter are
protected by supersymmetry so that a weak coupling computation can be continued
to strong coupling and compared in the AdS/CFT framework to the computation of the graviton
index in AdS space.
The matching of the two indices on the two sides corroborates the conjectured duality between them.
This is only one of the calculable exact results. By using localization \cite{Witten:1988hf},
we can in principle compute
the supersymmetric partition function (see 
\cite{Pestun:2007rz})
 on any manifold that preserves at least one complex supercharge
(or, in Euclidean space, two real supercharges), by reducing it to a matrix model, i.e. a finite
dimensional ordinary integral. 

Turning back to the case of the four-dimensional index, there are many available methods to obtain the matrix model formula 
for it. 
In \cite{Sen:1985dc,Sen:1985ph,Romelsberger:2005eg} the BPS states on the sphere have been found explicitly from the knowledge
of the spectrum of the Laplace operator: one needs to find the eigenmodes of the Laplace and Dirac
operator on the sphere and sum over all the modes.
Many of the bosonic modes will cancel out against the fermionic ones, and one finds that only the BPS
modes contribute to the index.
This is the most direct method, but it is in practice 
very difficult to
work out on generic supersymmetry preserving manifolds.
Another method is to compute the \emph{letter} index 
 \cite{Kinney:2005ej} for  
theories defined on a conformally flat background. 
In these cases, it is straightforward to obtain the  quantum numbers 
of the curved space fields 
because conformal mapping relates them to their flat space counterpart.
We can then identify the operators that  saturate the BPS inequality.
Nevertheless it is not simple to extend this method  to backgrounds
that are not conformally flat.
Finally, one can consider using localization. This amounts to picking a Q-exact term, generically related to
the supersymmetry transformations, and evaluate the ratio of two determinants, which
represents the full quantum corrections to the quantity one is considering \footnote{Recently the 
$\mathcal{N}=4$ superconformal index has been computed from localization in 
\cite{Nawata:2011un}.}.

Because of the difficulties of applying the previous methods to other manifolds,
it is simpler to identify just the BPS states in the Hilbert space.
One of the purposes of this paper is
to develop a method to achieve this aim.
The essential logic 
relies on the fact that the non-BPS modes are paired up by supersymmetry and hence the BPS modes
correspond to the kernel of the boson-fermion map.
The problem boils down to a set of first order differential equations.

 We also argue
a general relation between the BPS states 
and the set of
states that contribute non trivially to a corresponding partition function 
in one dimension less.
 More precisely, we will see that there exists a one-to-one
map between these two sets of states, and we identify the energy of each BPS state with
the quantum contribution of the dimensionally reduced state to the supersymmetric partition function. 
This relation has two immediate consequences. 
The first one is that
an index in $d$ dimensions reduces to a supersymmetric partition function for the
dimensionally reduced field theory, thus providing an argument 
which generalizes previous observations for the three-sphere
\cite{Dolan:2011rp,Gadde:2011ia,Imamura:2011uw}.

Another consequence  is the following. 
Since the states contributing to the partition function are the BPS states in one higher dimension,
we can uplift  the quantum contribution to
the partition function to the computation of the
energies of BPS states in one higher dimension and use the method
outlined above.
In this way, we only need to know the uplifted supersymmetry transformations and read from them
the pairing map. We believe that this  leads to  a simplification
in the computation of exact partition functions.
 
The paper is organized as follows. 
In section \ref{sec:4d} we review the definitions for the quantities
we are interested in, 
and explain our method to identify the BPS states in a general field theory.
We also describe in full detail the  relation between $d$-dimensional BPS states and the
$(d-1)$-dimensional physical states, focusing on the $4d/3d$ case for concreteness.
In section \ref{sec:example1} we show how the computations can be worked out for the
examples of the round and  squashed spheres. We give all the necessary details
to explicitly perform the computation, review previous results and discuss the physical
meaning of our results applied to the cases at hand. The reduction of these two
indices to the corresponding three-dimensional partition functions is shown in section \ref{sec:reduction}.
In section \ref{sec:extensions} we discuss generalizations of our technique to compute
the index to other manifolds and dimensions, while the idea to uplift the computation
of the partition function to a higher dimensional index is developed in section \ref{sec:uplift}.

\section{A correspondence between $4d$ and $3d$ states} \label{sec:4d}

One of the aims of the present paper is to develop a method to
identify the BPS states and to compute  
the 
 supersymmetric
index and the partition function on 
 a general class of manifolds.
In doing that, we will point out a connection between these two objects in
different dimensions. To be concrete, in this section we focus on the $4d/3d$
case.

Given a three-dimensional manifold ${\cal M}_3$ that
preserves some supersymmetry, and given a 
four dimensional supersymmetric theory defined on 
$X\equiv{\cal M}_3\times S^1$,
we can define two different quantities.
The first
one is the four dimensional 
superconformal index, defined on $X$,
that only takes contributions from BPS states.
It is the supersymmetric partition function
\be \label{eq:Isp}
{\cal I}_{sp}\left( t,y_i \right) = {\rm Tr} (-)^F \, e^{-\tau \Xi} \, t^H \prod_{i} y_i^{\gamma_i}
\ee
where $\Xi \equiv \{Q,Q^\dagger\}$, $F$ is the fermion number and the trace
is taken over every 
state
in the theory. $H$ and the $\gamma$'s form a
complete set of operators that commute
with the conserved supercharge $Q$.
In the following, we will call $H$
the energy operator and its eigenvalues the energies of the corresponding
eigenstates.
Moreover, the time direction is
identified with the circle and is thus periodic with period $\tau$.
The statement that the quantity \eqref{eq:Isp} only takes contributions
from BPS states means that for each bosonic state with $\Xi\neq 0$ there
exists a fermionic state with the same $(\Xi,H,\gamma_i)$ quantum
numbers; thus, the index turns out to be independent of $\tau$
due to the boson-fermion cancellations, and the trace can be taken over
the Hilbert space of $\Xi=0$ states.\footnote{By a "state" of the theory we
mean a configuration field which solves the equations of motion.} 
The index in (\ref{eq:Isp}) is the \emph{single particle} index. In the case of a gauge theory
one has to sum over all the possible gauge invariant configurations.

On the other
hand, we can reduce the given supersymmetric theory on ${\cal M}_3$ itself
and compute, at least in principle, the exact partition function for this theory
via localization. The latter reduces the partition function to the
matrix integral 
\be \label{eq:Ztot}
Z_{{\cal M}_3} \sim \int\!\! {\rm [d \sigma]} \,\, e^{-S_\ast} \frac{{\rm Pf}{D_F}}{\sqrt{\det{D_B}}}
\ee
where $[$d$\sigma ]$ represents the measure over the Cartan of the gauge group.
We have set the following notation for the two quantities we are interested
in. We denote by $S_\ast$ the classical action evaluated at the saddle
points, while the exact quantum contribution from the generic superfield $\Phi$ is
\be \label{eq:Zpert}
Z_{\Phi} = \frac{{\rm Pf}{D_F}}{\sqrt{\det{D_B}}}
\ee
where $D_F$ and $D_B$ are, respectively, a linear first order and
second order differential operator, and $\Phi$ labels both the chiral
and gauge multiplets.\footnote{The three-dimensional action may not be derived by dimensional
reduction of a corresponding four-dimensional theory. This is the case, for instance,
when a Chern-Simons term is present. 
The one loop determinants are not sensitive to these contributions and
our results also hold in those cases.}
A boson-fermion cancellation manifests itself
in the fact that some of the eigenvalues simplify between the
numerator and the denominator in \eqref{eq:Zpert}.

We argue that
the BPS states that contribute to
\eqref{eq:Isp} are in one-to-one correspondence to the  states contributing to 
\eqref{eq:Zpert}. 
More precisely, for each four-dimensional BPS  state with eigenvalue $E$
of $H$ there is a three-dimensional  state for which 
$E$ is an eigenvalue of the 
$D_B$  or of $D_F$   in the
case of boson or fermion respectively.

These states can be found
by solving a first order differential equation that can be directly read from
the supersymmetry transformations of the four dimensional theory. 
Finally, the saddle points in \eqref{eq:Ztot}
correspond to the zero energy states in the BPS spectrum: it
follows that, if there is no zero energy solution for a four-dimensional field $\Phi$,
the only  three-dimensional saddle point corresponds to $\Phi=0$.
 We will give more
details on this point in section \ref{sec:reduction}.

An argument for this correspondence
 is the following. It is well known that the
index \eqref{eq:Isp} does not depend on the radius of the compact time direction
and thus it does not change even when we shrink the circle to zero size.
More precisely, consider a fermionic state $\psi$ of a four-dimensional theory and
define a corresponding bosonic state
\be \label{BPSfermion}
\phi \equiv \zeta  \psi
\ee
where $\zeta$ is the Killing spinor which commutes with the BPS condition.
Then $\phi$ has the same $(\Xi,H,\gamma_i)$ quantum numbers and will
cancel the contribution of $\psi$ in \eqref{eq:Isp}, unless $\phi=0$ or,
equivalently, $\psi = \zeta  F$, with $F$ a scalar function
with the same $(\Xi,H,\gamma_i)$
quantum numbers of $\psi$. If $\psi$ is a state of the theory it satisfies 
the corresponding
 equation of motion: if we set $\psi \sim \psi_3(\vec x) e^{E t}$, it is thus easy to
recognize that the four-dimensional equation of motion can
be interpreted as the eigenvalue equation for a three-dimensional 
fermion with eigenvalue $E$. 

We now consider the bosonic states that contribute to the 
index:
we set up a map from the bosonic spectrum
to the fermionic one by
\be \label{BPSscalar}
\psi = i \sigma^\mu \tilde \zeta D_\mu \phi
\ee
which is  an infinitesimal supersymmetry transformation
(see below and section \ref{sec:curvedsusy}). We see
that every boson that contributes to the index is given by
$\sigma^\mu \tilde \zeta D_\mu \phi=0$. Once again, this can be interpreted
as an eigenvalue equation for a three-dimensional bosonic mode
 that contributes non trivially to the partition function.

The argument above can be cast in the following form. In four dimensions,
the supersymmetry transformations for the chiral multiplet are
\be \label{eq:susy1}
\begin{split}
\delta \phi &= \zeta  \psi \\
\delta \psi &= \zeta F + i \sigma^\mu \tilde \zeta \, D_\mu \phi  \\
\delta F &= \tilde \zeta \tilde \sigma^\mu {\cal D}_\mu \psi
\end{split}
\ee
where our conventions are explained in section \ref{sec:curvedsusy}. 
 Notice that the fermion equation of motion implies $\delta F=0$. 
This is a necessary condition that must be satisfied by the fermionic degrees of freedom.

The map that identifies the BPS states can be found to be
\bea \label{eq:tosolve2}
\begin{array}{rl}
\text{fermion:}& \quad \psi = \zeta F  \quad \text{and} \quad
\tilde \sigma^\mu {\cal D}_\mu \psi=0\\
\text{boson:}& \quad i \sigma^\mu \tilde \zeta \, D_\mu \phi = 0\\
\end{array}
\eea
We further notice the following. The system \eqref{eq:tosolve2}  implies
$|\delta\psi|^2=0$ (in the absence of F-terms),
 and when the fields are  independent  on the time
direction, we can dimensionally reduce the latter equation which becomes
the three-dimensional saddle point equation used in the localization setting.

We now turn to the vector multiplet. Once again we can set up a map
between the bosonic and the fermionic Hilbert space by using the supersymmetry
transformations. Analogously to the discussion above, all the contributions will
cancel out but those coming from the zero modes of the map.

In four dimensions, the physical fields in the vector multiplet are a gauge field $v_\mu$ and
the gaugino $\lambda$. The supersymmetry transformations are
\bea \label{eq:susy2}
\begin{split}
\delta v_\mu &= i \zeta \sigma_\mu \tilde\lambda 
       - i \tilde \zeta \tilde \sigma_\mu \lambda \\
\delta \lambda &= \zeta D + i \sigma^{\mu\nu} F_{\mu\nu} \zeta \\
\delta D &= i \tilde \zeta \tilde \sigma^\mu {\cal D}_\mu \lambda
       - i \zeta \sigma^\mu {\cal D}_\mu \tilde \lambda
\end{split}
\eea
The map that identifies BPS states
can be found to be
\bea \label{eq:susy2bis}
\begin{array}{rl}
\text{gaugino:} &~~~ 
-i \tilde\zeta \tilde\sigma_\mu \lambda = \partial_\mu \varphi
\quad \text{and} \quad 
i \tilde \sigma^\mu {\cal D}_\mu \lambda = 0
\\
\text{gauge boson:}&~~~ i \sigma^{\mu\nu} F_{\mu\nu} \zeta = 0 \\
\end{array}
\eea
 where once again $\delta D=0$ is a necessary condition for the
gaugino degrees of freedom.

In the first line we had set the gauge field to a pure gauge configuration
because
any such solution
 does not give rise to a state in the Hilbert space
of the theory and hence the gaugino does not have a superpartner state.
Alternatively, we
could have considered the map between the field strength $F_{\mu\nu}$ and the
gaugino, which leads to the same condition.
It is easy to see that the zero energy solutions to \eqref{eq:susy2} reduce to the 
three-dimensional saddle point equations for a three-dimensional Q-exact
action. The set of non-trivial solutions for $\lambda$ and $F_{\mu\nu}$ gives
the Hilbert space we have to trace over in equation \eqref{eq:Isp}, or alternatively
the spectrum of eigenvalues contributing to \eqref{eq:Zpert}.

To summarize, we are led to the conclusion that  a priori
different exact results in different dimensions are related one to the other.
The reduction of the four-dimensional index to the three-dimensional
partition function follows directly from the proposed connection between the 
four-dimensional  and three-dimensional states.
While we will give more details on this point
in section \ref{sec:reduction}, we stress here that our claim is stronger than
the dimensional
reduction of the superconformal index to the partition function,  because we set up
a one-to-one map between states and eigenvalues of different operators.

On the one hand we look for eigenstates of the four-dimensional Hamiltonian, 
on the other hand we look for eigenstates of the equations of motion derived 
from a Q-exact three-dimensional Lagrangian,
that contributes to the  partition function.
While the former is a first order differential operator, the latter is in general a
second order one.

In the next section we will explicitly check our proposal in two cases: ${\cal M}_3=S^3$,
in which case we can compare with known results, and ${\cal M}_3=S^3_b$,
with $S^3_b$ a squashed sphere. In the latter case, because the index
is a topological invariant, it can be cast in the same form as the index on a sphere
via a redefinition of its arguments. However, we show that one can keep the
original definitions and define a natural limit to recover the three-dimensional
partition function on the squashed three-sphere computed in \cite{Hama:2011ea}.
We thus conclude that, although the
index does not carry different physical information on different but topologically
equivalent manifolds, it contains different information when we reduce the
four-dimensional theory to a three-dimensional one by shrinking the time circle.
It thus becomes interesting, from a three-dimensional point of view, to compute
the four-dimensional index even on topologically equivalent manifolds.

\section{Examples: sphere and squashed spheres}
\label{sec:example1}

\subsection{Review of rigid supersymmetry on a curved manifold} \label{sec:curvedsusy}

We review here a simple and recent procedure
to place an $\mathcal{N}=1$ supersymmetric theory on a curved four-dimensional manifold \cite{Festuccia:2011ws}.
The basic idea is to start with $\mathcal{N}=1$ supergravity and take an appropriate
limit such as to decouple gravity but preserve the classical background
configuration. Because a convenient off-shell formulation and its couplings
to matter fields are known, the gravitino supersymmetry transformation looks
very simple \cite{Sohnius:1981tp,Sohnius:1982fw}
\be \label{eq:gravitino}
\begin{split}
\delta \psi_\mu &=  -2 {\cal D}_\mu \zeta - 2 i V^\nu \sigma_{\mu\nu} \zeta\\
\delta \tilde\psi_\mu &=  -2 {\cal D}_\mu \tilde\zeta + 2 i V^\nu \tilde\sigma_{\mu\nu} \tilde\zeta\\
D_\mu &\equiv \nabla_\mu - i q_A A_\mu \\
{\cal D}_\mu &\equiv D_\mu - i q_V V_\mu
\end{split}
\ee
where $q_A$ and $q_V$ are the charges (
under the $A$ and $V$ background gauge 
fields)  of the field  on which the covariant derivative is acting on. For the Killing spinor $\zeta$,
$q_A^\zeta=1$ and $q_V^\zeta=-1$, and $\tilde\zeta$ has opposite quantum numbers.
Because gravity is decoupled,
one can give an expectation value to
the background gauge fields $A$ and $V$ and to the metric without
having to take care of their equations of motion.

Once we have found a solution to $\delta \psi_\mu=0$ and $\delta \tilde\psi_\mu=0$, the supersymmetry
 transformations of the matter fields are
\be \label{eq:susy}
\begin{split}
\begin{array}{lll}
\delta \phi = \zeta  \psi  & \qquad q_A^\phi=q & \qquad q_V^\phi = -1/2 \\
\delta \psi = \zeta F + i \sigma^\mu \tilde \zeta \, D_\mu \phi & \qquad q_A^\psi=q-1 & \qquad q_V^\psi = 1/2  \\
\delta F = \tilde \zeta \tilde \sigma^\mu {\cal D}_\mu \psi & \qquad q_A^F=q-2 & \qquad q_V^F = 3/2
\end{array}
\end{split}
\ee
for the chiral multiplet, and, in the Wess-Zumino gauge,
\be \label{eq:susyvec}
\begin{split}
\begin{array}{lll}
\delta v_\mu = i \zeta \sigma_\mu \tilde\lambda 
       - i \tilde \zeta \tilde \sigma_\mu \lambda 
        & \qquad q_A^v=0 & \qquad q_V^v = -1/2 \\
\delta \lambda = \zeta D + i \sigma^{\mu\nu} F_{\mu\nu} \zeta
       & \qquad q_A^\lambda=1 & \qquad q_V^\lambda = -3/2  \\
\delta D = i \tilde \zeta \tilde \sigma^\mu {\cal D}_\mu \lambda
       - i \zeta \sigma^\mu {\cal D}_\mu \tilde \lambda  
         & \qquad q_A^D=0 & \qquad q_V^D = -1/2
\end{array}
\end{split}
\ee
for the vector multiplet. An action which is invariant under these supersymmetry
transformations is\footnote{We are considering Euclidean signature.
The derivatives should be understood to be covariant with respect to the gauge field too,
but due to the invariance of the index under continuous transformations, we can
switch off the gauge coupling without changing the result. \label{fn:gauge}}
\bea
{\cal L} &=& {\cal L}^B + {\cal L}^F \\
\frac{{\cal L}^{B}}{\sqrt{g}}  &=&\left(-\frac{1}{4} {\cal R} - \frac{3}{2}  V_{\mu}V^{\mu}\right) 
   q \phi \bar\phi  -D_{\mu}\phi D^{\mu}{\bar \phi} + F \bar F \nonumber \\
   & & + {iV^{\mu}}\left(\bar\phi D_{\mu} \phi- \phi D_{\mu} \bar \phi \right) 
    +\frac{1}{4} F_{\mu\nu} F^{\mu\nu} + \frac{1}{2} D^2 \nonumber \\
\frac{{\cal L}^{F}}{\sqrt{g}}  &=&    -{i} \tilde \lambda \tilde \sigma^\mu {\cal D}_\mu \lambda
    -{i} \tilde \psi \tilde \sigma^\mu {\cal D}_\mu \psi \nonumber
\eea

\subsection{Supersymmetry on a general squashed sphere}

In this section we present all the necessary results to work out the examples of the sphere
and the squashed sphere to be described in full details in the next sections.
We give the full expressions in the case of the squashed sphere, while supersymmetry
on the sphere is recovered by taking an appropriate limit. 
Some of the results shown here can be also found in \cite{Dumitrescu:2012ha,Klare:2012gn}.

The squashed sphere $S^3_b$ enjoys a $U(1)^2$ isometry. The latter is
made manifest if we choose the Hopf coordinates
$x^\mu = \{t,\theta,\alpha,\beta\}$, with $\mu=1,\ldots,4$, such as $t$ denotes the
Euclidean time
coordinate compactified on a circle. The coordinates $\alpha$ and $\beta$ have range
$[0,2\pi)$ while $\theta \in [0,\pi/2]$. The metric reads
\be
\label{metricgeneric}
ds^2=  dt^2 + f(\theta)^2 d \theta^2+ a^2 \cos(\theta)^2 d\alpha^2 + b^2 \sin(\theta)^2 d\beta^2
\ee
where  $f(\theta)$ is regular on $(0,\pi/2)$ and  
 and $f(0)=b$ and $f(\pi/2)=a$. Moreover  the manifold even if compact can
also be locally hyperbolic. 
 The Ricci tensor is
\begin{equation}
{\cal R}=\frac{6 f(\theta) + 4 \cot(2 \theta) f^\prime(\theta)}{f(\theta)^3}
\end{equation}
In principle, we could have introduced two parameters, say $R_1$ and $R_2$,
multiplying the time and squashed sphere
terms respectively in the metric. The gravitino variation
then imposes $R_1=R_2$, and the overall factor can be set to unity by a redefinition
of the time period, which does not affect our computations.

The Killing spinor equations in the new minimal formalism are solved by
\bea \label{eq:killing}
\begin{split}
\zeta_\alpha &= -\frac{i}{\sqrt{2}} e^{\frac{i}{2} \left( \alpha+\beta \right)} \left(
\begin{array}{c}
e^{-\frac{i}{2} \theta} \\
i e^{\frac{i}{2} \theta}
\end{array}
\right) \qquad \qquad
\tilde \zeta^{\dot\alpha} = -\frac{i}{\sqrt{2}} e^{-\frac{i}{2} \left( \alpha+\beta \right)} \left(
\begin{array}{c}
e^{\frac{i}{2} \theta} \\
i e^{-\frac{i}{2} \theta}
\end{array}
\right) \\
V_\mu dx^\mu &=  -\frac{i}{f(\theta)} dt \\
A_\mu dx^\mu &= -\frac{i}{f(\theta)} dt + \left( \frac{1}{2}-\frac{a}{2 f(\theta)} \right) d\alpha + \left( \frac{1}{2}-\frac{b}{2 f(\theta)} \right) d\beta
\end{split}
\eea
which shows that, for generic squashing parameters $a,b$ there are two supercharges.
In the round sphere limit we can find two more Killing spinors, showing that the manifold
enjoys four supercharges. Our results only rely on the existence of two real
supercharges, and we choose \eqref{eq:killing}  which is a convenient choice both
for the sphere and the squashed spheres.

With our choice of background fields, the algebra involving the two supercharges above is
\bea \label{eq:algebra}
\begin{split}
[H,Q]&=0 \qquad \qquad  \{Q,\tilde Q\}=H-\frac{R}{2}\left(\frac{1}{a}+\frac{1}{b}\right) + 2 J_3 \\
[R,Q]&=-Q \qquad \qquad \left[ 2 \tilde J_3 + \frac{R}{2} \left(\frac{1}{a}-\frac{1}{b} \right), Q \right] = 0 \\
H&\equiv \partial_t \qquad\qquad 2 J_3 \equiv -\frac{i}{a} \partial_\alpha-\frac{i}{b} \partial_\beta
 \qquad \qquad 2 \tilde J_3 \equiv \frac{i}{a} \partial_\alpha-\frac{i}{b} \partial_\beta
\end{split}
\eea

From the supersymmetric action
\bea
\frac{{\cal L}}{\sqrt{g}}  &=&\left(-\frac{1}{4} {\cal R} + \frac{3}{2 f^2} \right) 
   q \phi \bar\phi - D_{\mu}\phi D^{\mu}{\bar \phi} + F \bar F \nonumber \\
   & & + \frac{1}{f} \left( \bar\phi D_t \phi- \phi D_t \bar \phi \right) 
    + \frac{1}{4} F_{\mu\nu} F^{\mu\nu} + \frac{1}{2} D^2 \nonumber \\
  & &  - i \tilde \lambda \tilde \sigma^\mu {\cal D}_\mu \lambda
    - i \tilde \psi \tilde \sigma^\mu {\cal D}_\mu \psi
\eea
we can derive the following equations of motion
\bea
\begin{split}
\Delta_\phi \phi & \equiv \left( D^\mu D_\mu + \frac{2}{f} D_t + q \left(-\frac{1}{4} {\cal R} + \frac{3}{2 f^2}
    \right) \right) \phi = 0 \\
\Delta_\psi \psi & \equiv  i \tilde \sigma^\mu {\cal D}_\mu \psi = 0 \\
\nabla_\mu F^{\mu\nu} &= J^\nu \\
\Delta_\lambda \lambda & \equiv  i \tilde \sigma^\mu {\cal D}_\mu \lambda = 0
\end{split}
\eea
where $J^\nu$ is an appropriate current which vanishes in the $g_{YM}\to 0$ limit.

\subsection{The three sphere}

In this section we apply the proposal explained above to the 
calculation of the superconformal index on $S^3\times S^1$, and show that
it agrees with previous results \cite{Romelsberger:2005eg,Kinney:2005ej}.
We start by reviewing the calculation of the 
index in terms of the expansion of the field
configurations in spherical harmonics.
There are two multiplets contributing to the index, the chiral 
multiplet $\Phi=(\phi,\psi)$ with $R[\phi]=q$ and the vector multiplet $\mathcal{V}=(v,\lambda)$. 

The harmonic expansion has first been done in \cite{Sen:1985dc,Sen:1985ph} and we report it here
with conventions adapted to Euclidean signature. The algebra chosen there coincides
with the round sphere limit of our equation \eqref{eq:algebra}, so the definition of the index
works without further changes.

The eigenvalues of the Laplace operator acting on scalars on the three-sphere are
$-j(j+2)$, with $j$ a nonnegative integer.
 By plugging the expansion scalar field 
   \begin{equation}
\label{eq:bosmodes}
 \Phi = \sum_n  a_n \Phi_+(n) e^{E_+{(n)} t} + c_n^{\dagger} \Phi_-^\dagger (n) e^{ E_-{(n)} t}
 \end{equation}
in the equation of motion, one sees that, including the R-charge contribution, the normal modes are
  \begin{center}
  \begin{tabular}{c||cc}
Wave function  &$E$ & $(J_3,\tilde J_3)$\\
\hline
 $a_n$ &$j+ q$&$\left(\frac{j}{2},\frac{j}{2} \right)$\\
   $c_n^{\dagger}$ &$-j-2+q$&$\left(\frac{j}{2},\frac{j}{2} \right)$
  \end{tabular}
  \end{center}
  where $j\geq 0$ and in the last column we have indicated the representation of the fields
under the Cartan subgroup of the isometry group of the sphere. 
A field is in the
$\left(j/2,j/2\right)$ representation means that the $j_3$ and $\tilde j_3$ eigenvalues
can range from $-j/2$ to $j/2$ at fixed $j$.

An analogous expansion holds for the chiral fermion
\begin{equation}
\label{eq:fermmodes}
\Psi=\sum_n b_n \Psi_{+}(n) e^{ E_+{(n)} t} + d_n^{\dagger} \Psi_{-}^\dagger(n)  e^{ E_-{(n)} t} 
 \end{equation}
which gives
  \begin{center}
  \begin{tabular}{c||cc}
Wave function   & $E$ & $(J_3,\widetilde{ J_3})$\\
\hline
 $b_n$ &$j+q$&$\left(\frac{j-1}{2},\frac{j}{2} \right)$\\
   $d_n^{\dagger}$ &$-j+ q-1$&$\left(\frac{j}{2},\frac{j-1}{2} \right)$
  \end{tabular}
  \end{center} 
where $-j(j+1)$, $j\geq 1$, are the eigenvalues of the Laplace operator on spinors on the
three-sphere.

To compute the index, one has in principle to sum over all these states. However, we know
that the index only takes contributions from BPS states, i.e. states that satisfy $\Xi=0$.
It is easy to realize that this constraint fixes the $j_3=-j/2$ particle state for the scalar
field and the $j_3=j/2$ antiparticle state for the fermion, while $\tilde j_3$ is unconstrained
because it does not appear in $\Xi$.
By summing over all these states
the contribution to the  superconformal  index  of
the chiral multiplet is 
\begin{equation}
I_{\Phi} =  \sum_{j,\widetilde {j_3}} (-1)^F e^{- \tau \Xi} t^{H } y^{2 \tilde j_3}=
\frac{t^q-t^{2-q}}{(1-t y)(1-t/y)} 
\label{equationindexsphere}
\end{equation}

\begin{figure}
\begin{center}
  \includegraphics[width=6cm]{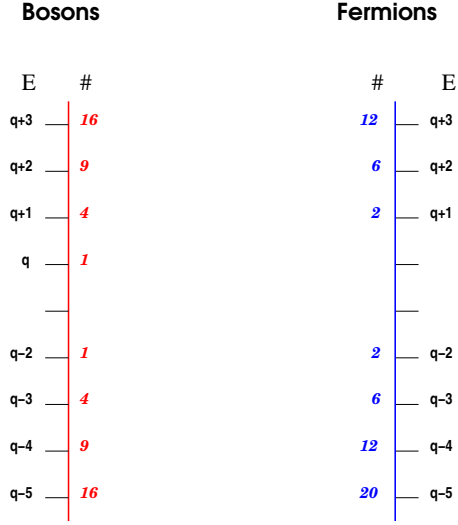}
\caption{A schematic  structure of the pairings among the modes. Here $E$ is the energy of the mode and $\#$  is
the number of bosonic and fermionic modes with a given energy.}
\label{Pairingmodes}
\end{center}
\end{figure}

This structure of pairing and un-pairing among the modes is  explicitly shown in the Figure
\ref{Pairingmodes}. In general, for $j\geq 0$,  we have the following structure
\begin{center}
\begin{tabular}{c||cc|cc}
&$E_+$&Degeneration&$E_-$&Degeneration\\
\hline
Boson&$j+q$&$(j+1)^2$&$-j-2+q$&$(j+1)^2$\\
Fermion&$j+q$&$j(j+1)$&$-j-2+q$&$(j+1)(j+2)$\\
\end{tabular}
\end{center}
The BPS modes are the modes unpaired in this table, and they
are counted by the superconformal index as explained above.

We can repeat the above procedure for the vector multiplet. In the case of the gaugino we have
  \begin{center}
  \begin{tabular}{c||cc}
Wave function  &$E$ & $(J_3,\tilde J_3)$\\
\hline
 $b_n$ &$j$&$\left(\frac{j-1}{2},\frac{j}{2} \right)$\\
   $d_n^{\dagger}$ &$-j-1$&$\left(\frac{j}{2},\frac{j-1}{2} \right)$
  \end{tabular}
  \end{center}
  with $j\geq 1$.
For the vector field one can expand in terms of the spin-1 spherical harmonics and the modes are
  \begin{center}
  \begin{tabular}{c||cc}
Wave function  &$E$ & $(J_3,\tilde J_3)$\\
\hline
 $a_n$ &$j+1$&$\left(\frac{j-1}{2},\frac{j+1}{2} \right)$\\
   $c_n^{\dagger}$ &$-j-1$&$\left(\frac{j+1}{2},\frac{j-1}{2} \right)$
  \end{tabular}
  \end{center}
    with $j\geq 1$.
    By summing over all these states
the contribution of the vector multiplet  to the superconformal  index  is 
\begin{equation}
 I_{V} =  \sum_{j,j_3\widetilde {j_3}}(-1)^F e^{- \tau \Xi} t^{H } y^{2 \tilde j_3}=
\frac{2t^2- t (y+1/y)}{(1-t y)(1-t/y)}
\end{equation}

In the rest of this section we apply our prescription to obtain the BPS states in a 
different way, in which it is not necessary to solve for the whole spectrum.
We start by considering the metric as in (\ref{metricgeneric})
with $a=b=1$. The two angles $\alpha$ and $\beta$ can be associated to the
Cartan subgroup of the $SU(2)^2$ isometry group of the metric.

We start by solving the equation (\ref{eq:tosolve2}) for the BPS fermion in the chiral multiplet .
Once we write the fermion as  $\psi= \zeta F$ and solve the equation $\Delta_{\psi} \left(\zeta F\right)=0$,
 we expand $F$ as $F = e^{E t+i n \alpha  + i m \beta } g_{\psi}(\theta)$, where
 $E$ is the eigenvalue associated to the $S^1$ and  $n$ and $m$
are integer numbers associated to the two $SU(2)$ in the $S^3$, parameterized by the periodic
coordinates $\alpha$ and $\beta$ in the metric. We obtain
\begin{eqnarray}
\left\{
\begin{array}{ccc}
g_{\psi}'+i \, g_\psi ( 2+E+m+n-q+i\, m \cot \theta -i\, n  \tan \theta)&=&0\\
g_\psi'-i g_\psi ( 2+E+m+n-q-i\,m \cot \theta +i\, n  \tan \theta)&=&0\\
\end{array}
\right.
\end{eqnarray}
These two equation can be simultaneously solved for 
$E =q-2-m-n$
 and the solution is
 \begin{equation}
 g_\psi(\theta) =  \sin^m \theta \cos^n\theta \qquad\qquad \mbox{for} \qquad \theta\sim 0, \pi/2
 \end{equation}
 that is square integrable if  $m,n\ge 0$ .
This represents the contributions of the BPS fermion to the index. 
 Because $E$ is negative, we have found that the corresponding state is an 
\emph{antiparticle} mode of the fermion. Thus, when we plug its quantum numbers in the index,
we have to flip their signs: 
the energy of the field is $E^{(\psi)}=-E=2-q+m+n$. The other operator that commutes 
 with the supercharge is
 $\tilde J_3$ that has eigenvalues 
$m-n$. The fermionic contribution to the index is then
\begin{equation}
I_\psi= - \sum_{m,n\ge 0} t^{n+m+2-q} y^{m-n} = -\frac{t^{2-q}}{(1-t y)(1-t/y)}
\end{equation}
We parameterize the  BPS   boson   
as $\phi = e^{E t+i n \alpha  + i m \beta} g_{\phi}(\theta)$ 
and the equation  (\ref{BPSscalar}) becomes
\begin{eqnarray}
\left\{
\begin{array}{ccc}
g_{\phi}'+i g_{\phi}(E+m+n-q-i\, m \cot  \theta +i \,n \tan  \theta )&=&0\\
g_{\phi}'-i g_{\phi} ( E+m+n-q+i\, m \cot  \theta +i \,n \tan  \theta )&=&0\\
\end{array}
\right.
\end{eqnarray}
The two equations are compatible if 
$E = q-m-n$ and the solution is 
 \begin{equation}
 g_\phi(\theta) =  \frac{1}{\sin^m \theta \cos^n\theta} \qquad\qquad \mbox{for} \qquad \theta\sim 0, \pi/2
 \end{equation}
and square integrability imposes  $m,n \le 0$.
 The  BPS boson that contributes to the index is the particle in the expansion in terms of 
 creation and annihilation operators, with energy $E^{(\phi)}=E$.
 The bosonic index is
 \begin{equation}
I_\phi = \sum_{m,n\ge 0} t^{n+m+q} y^{m-n} = \frac{t^{q}}{(1-t y)(1-t/y)}
\end{equation}

We now turn to the vector multiplet. 
In the case of the gaugino we read the pairing map from the transformation of $F_{\mu \nu}$.
The BPS modes are the solution of the equation 
\begin{equation}
\partial_{\mu}( \tilde \zeta \bar \sigma_{\nu} \lambda)   -
 \partial_{\nu}( \tilde \zeta \bar \sigma_{\mu} \lambda)   = 0
\end{equation}
This equation is solved by 
\begin{equation}
\label{solgaugino}
 \tilde \zeta \bar \sigma_{\mu} \lambda = \partial_{\mu} \Phi
 \end{equation}
Alternatively, one can require that the SUSY variation for $v_\mu$ gives a purely longitudinal field.
 We then impose the usual ansatz dictated by the $U(1)$ symmetries 
\begin{equation}
\Phi = e^{E t + i \alpha n + i \beta m} h_{\Phi} (\theta)
\quad,
\quad
\lambda = e^{E t + i \alpha\left(n+\frac{1}{2}\right) + i \beta \left(m+\frac{1}{2}\right)}
\left(
\begin{array}{c}
\lambda_{1}(\theta)\\
\lambda_2 (\theta) 
\end{array}
\right)
\end{equation}  
and we plug it in (\ref{solgaugino}). 
Moreover we impose that $\lambda$ satisfies its equations of motion.
In this way we find 
\begin{eqnarray}
\label{gaugionosolns}
E &=& -m-n \nonumber \\
\lambda_1(\theta)& =& \frac{1}{\sqrt 2}e^{\frac{i \theta }{2}} h_\Phi(\theta) \left(\frac{ m}{ \sin \theta}+i\frac{ n}{\cos \theta}\right) \nonumber
\\
\lambda_2(\theta) &=&-\frac{1}{\sqrt 2}e^{-\frac{i \theta }{2}}h_\Phi(\theta) \left(i \frac{ m}{ \sin \theta}+\frac{ n}{\cos \theta}\right) \nonumber \\
h_\Phi'(\theta) &=& h_\Phi(\theta) \left(n\tan\theta-m \cot \theta \right)
\end{eqnarray}
The equation for $h_\Phi(\theta)$ tells us that the solution is square integrable for $m,n\leq 0$,
but we exclude the vanishing solution corresponding to $(m,n)=(0,0)$.
Thus the energy is positive and the gaugino contribution
to the index is  
\begin{equation}
   I_\lambda=-\left(
\sum_{m,n=-\infty}^{0}   t^{-m-n}y^{m-n} - 1 \right)
=\frac{t^2 - t (y+1/y)}{(1-t y)(1-t/y)}
\end{equation}
where the second term comes from subtracting the  $(m,n)=(0,0)$ contribution.

The gauge field works as follows. 
First we impose that the BPS equation is satisfied
\begin{equation} \label{eq:BPSgauge}
\sigma^{\mu \nu} \zeta F_{\mu \nu} = 0 
\end{equation}
We consider the Abelian case and define the components of the EM field as
$\mathcal{E}_i = F_{t i}$ and $2 \mathcal{B}_i = \epsilon_{ijk} F_{jk}$ where the latin letters  label the $S^3$ coordinates.
We parametrize these fields with the ansatz
\begin{equation}
\mathcal{E}_i(t,\theta,\alpha,\beta) = e^{E t + i \alpha n + i \beta m} \mathcal{E}_i (\theta)
\quad
,\quad
\mathcal{B}_i(t,\theta,\alpha,\beta) = e^{E t + i \alpha n + i \beta m} \mathcal{B}_i (\theta)
\end{equation}
From \eqref{eq:BPSgauge} we derive
 the following three equations
\begin{eqnarray}
\begin{split}
\csc  \theta  \sec  \theta \mathcal{B}_{\theta }+\mathcal{E}_{\theta }&\equiv y(\theta )\\
\csc  \theta  \mathcal{B}_{\alpha }+\sec  \theta  \mathcal{E}_{\alpha }&= -i \sin  \theta \,  y(\theta )\\
\sec  \theta  \mathcal{B}_{\beta }+\csc  \theta  \mathcal{E}_{\beta }&= i \cos  \theta \, y(\theta )
\end{split}
\end{eqnarray}
where $y(\theta)$ is arbitrary.
The other equations are the Maxwell equation (or equivalently the Bianchi identities and the equations of motion).
The equations of motion $\mathcal{D}_{\mu}F^{\mu \nu}=0$ are
 \begin{eqnarray}
\begin{split}
  -i \left(m \sec ^2\theta \mathcal{E}_{\alpha }+n \csc ^2\theta
    \mathcal{E}_{\beta }\right)+E_{\theta } -2 \cot 2 \theta
  \mathcal{E}_{\theta }-\mathcal{E}_{\theta }' &= 0
  \\
  i n \csc ^2\theta \mathcal{B}_{\alpha }-i m \sec^2\theta
  \mathcal{B}_{\beta }+E \, \mathcal{E}_{\theta }&= 0
  \\
  -i n \csc ^2\theta f^2 \mathcal{B}_{\theta }+\tan \theta
  \mathcal{B}_{\beta }+E \, \mathcal{E}_{\alpha }-\mathcal{B}_{\beta
  }+\mathcal{B}_{\beta }'&= 0
  \\
  -i m \sec ^2\theta \mathcal{B}_{\theta }+\cot \theta
  \mathcal{B}_{\alpha }+E \, \mathcal{E}_{\beta }+\mathcal{B}_{\alpha
  }\mathcal{B}_{\alpha }'&= 0
\end{split}
\end{eqnarray}
and the Bianchi identities $\partial_{ [\mu} F_{\nu\rho]}=0$ are
\begin{eqnarray}
\begin{split}
i m \mathcal{B}_{\alpha }+i n \mathcal{B}_{\beta }+\mathcal{B}_{\theta }'&= 0
\\ 
E \mathcal{B}_{\beta }+i m \mathcal{E}_{\theta }-\mathcal{E}_{\alpha }'&= 0
 \\
E \mathcal{B}_{\alpha }+i n \mathcal{E}_{\theta }-\mathcal{E}_{\beta }'&= 0
 \\
E \mathcal{B}_{\theta }+i n \mathcal{E}_{\alpha }-i m \mathcal{E}_{\beta }&= 0
\end{split}
\end{eqnarray}
We then have eleven equations for seven variables (the energy and the non zero components of the electromagnetic fields). 
Even if the system looks overdetermined these equations are linearly dependent. By expressing every function in terms of 
$y(\theta)$ 
and $E$  we obtain
\begin{equation}
\left(\partial_{\theta} -(m-1)\cot \theta +(n-1) \tan\theta \right)y(\theta) = 0 
\quad,
\quad
E = -m - n
\end{equation}
The solution is square integrable for $m,n\geq 1$. In this case the contribution comes from the antiparticle
in the mode expansion  and the index is
\begin{equation}
 I_{B}^{(V)}=
 \sum_{m=1}^{\infty}\sum_{n=1}^{\infty} t^{m+n} y^{n-m}=\frac{t^2}{(1-t y)(1-t/y)}
\end{equation}
If we consider a non abelian gauge group we must  add an extra  chemical potential for the 
gauge symmetry. Indeed since the index is a topological invariant the gauge coupling does not play any role and we only need to take care 
of the fact that the vector multiplet transforms in the adjoint representation.
The gauge invariant combinations are given by the Plethystic exponential 
after  integrating over the Haar measure
\cite{Aharony:2003sx,Benvenuti:2006qr}.

\subsection{Squashed spheres}

The superconformal index on the squashed sphere is expected to coincide 
with the one computed in the round limit, up to a redefinition of the variables.
Indeed this manifold preserves the topological properties of
$S^3$ and  this guarantees that the index 
does not change under squashing.

This can be shown with a simple argument based on the definition of
the index.  Indeed the index on the three sphere is defined as \bea
{\cal I}_{S^3\times S^1} (t,y) &=& \text{Tr}(-1)^F e^{- \beta \{
  Q,Q^\dagger\}} t^{H} y^{2 \tilde J_3} =
\sum_{BPS} \text{Tr}(-1)^F t^{R-2 \tilde J_3} y^{2 \tilde J_3} \nonumber \\
&=& \sum_{BPS} \text{Tr}(-1)^F t^{R+J_\alpha+J_\beta}
y^{J_\alpha-J_\beta} \eea where $J_{\alpha}\equiv i\partial_\alpha$
and $J_{\beta}\equiv i\partial_\alpha$ are the generators of the two
$U(1)$'s in the Hopf fibration.  By defining $p=t y $ and $q=t/y$ the
index becomes (this change of coordinates has been first considered in
\cite{Dolan:2008qi})
\begin{equation} \label{indexpq}
{\cal I}_{S^3\times S^1} (p,q) =
  \sum_{BPS} \text{Tr}(-1)^F p^{R/2+J_\alpha} q^{R/2+J_\beta}
\end{equation}
The same definition of the index on the squashed sphere is \bea
{\cal I}_{S^3\times S^1} (t,y) &=& \text{Tr}(-1)^F e^{- \beta \{ Q,Q^\dagger\}} t^{H} y^{2 \tilde J_3+\frac{R}{2}\left(\frac{1}{a}-\frac{1}{b}\right)} \nonumber \\
&=& \sum_{BPS} \text{Tr}(-1)^F
t^{\frac{R}{2}\left(\frac{1}{a}+\frac{1}{b}\right)
  +\frac{J_\alpha}{a}+\frac{J_\beta}{b}}
y^{\frac{J_\alpha}{a}-\frac{J_\beta}{b}+\frac{R}{2}\left(\frac{1}{a}-\frac{1}{b}\right)}
\eea By defining $p=(t y)^{\frac{1}{a}} $ and $q=(t /y)^{\frac{1}{b}}$
the index on the squashed sphere is defined as (\ref{indexpq}) and its
definition coincides with the one for the round case as expected.
Then the index is expected to coincide because the two spaces have the
same topology, and the same BPS states contributing to the index in
the round case contribute to the index in the squashed case.

In this section we explicitly show this result by exploiting the power of our prescription for the identification of the BPS states.
Indeed there are no known result for expansion in terms of harmonics on these spaces
and a direct calculation is not at hand.

We start by writing the fermion in the chiral multiplet as $\psi= \zeta F$ and solve the equation $\Delta_{\psi} \left(\zeta F\right)=0$,
where we expand $F$ as $F = e^{E t+i n \alpha + i m \beta } g_{\psi}(\theta)$, obtaining the following set
\begin{eqnarray}
\left\{
\begin{array}{l}
 g_{\psi}'
=-
\frac{e^{-i \theta } g_{\psi}}{2 a b}
\left(
2 a b e^{i \theta } (i E f-(q-2) \cot{2 \theta})+ 
f \left(
\frac{(a (q-2(m+1))}{\sin\theta} 
+
\frac{i b (2(n+1)-q)}{\cos\theta}
\right)
\right) \nonumber \\
g_{\psi}'=\frac{g_{\psi}}{2 a b} 
\left(\frac{2 a b (q - 2)}{ \tan 2 \theta} +f \left(\frac{a e^{i \theta } (2 (m + 1) - q) }{\sin\theta} +\frac{i b (2 a E + (2(n+1)-q)(1+i \tan \theta))}{\cos \theta}\right)
\right) \nonumber \\
\end{array}
\right.
\end{eqnarray} 
 These two equations can be simultaneously solved if
 \begin{equation}
E= \frac{q}{2}\left(\frac{1}{a}+\frac{1}{b} \right) -\frac{1}{a}-\frac{1}{b}  - \frac{n}{a}  - \frac{m}{b}  
 \end{equation}
Square integrability requires the quantum numbers $m,n\geq 0$ as in the case of the sphere.
  The mode contributing to the index is an \emph{antiparticle} and its energy is
$E_{\psi} = -E$. By summing over the BPS states we have 
\begin{equation}
I_\psi=-\frac{(t/y)^{\frac{2-q}{2 b}} (t y)^{\frac{2-q}{2 a}}}{\left(1-(t/y)^{\frac{1}{b}}\right) \left(1-(t y)^{\frac{1}{a}}\right)}
\end{equation}

The equations for the scalar  $\phi = e^{E t+i \alpha m + i \beta m} g_{\phi}(\theta)$ become
\begin{eqnarray}
\left\{
\begin{array}{l}
g_{\phi}'=-\frac{g_{\phi} }{2 a b}\left( \frac{2 a b q }{\tan 2 \theta}+f \left(2 i a b E+e^{i \theta } \left( \frac{a (2 m-q)}{\sin\theta }+\frac{i b (2 n-q)}{\cos\theta} \right)\right)\right)  \\
g_{\phi} '=-\frac{g_{\phi} e^{-i \theta }}{2 a b}
\left(
\frac{2 a b q e^{i \theta }}{\tan 2 \theta}
+f 
\left(
\frac{a (2 m-q)}{\sin\theta} -i b \left(2 a e^{i \theta } E +\frac{ (2 n-q)}{\cos\theta} \right)\right)\right)
\end{array}
\right.
 \end{eqnarray}
 They can be simultaneously solved if
 \begin{equation}
E=\frac{q}{2}\left(\frac{1}{a}+\frac{1}{b}\right) -\frac{n}{a}-\frac{m}{b}
 \end{equation}
 with $m,n \leq 0$. This constraint fixes $E_{\phi}=E$ and the index for the scalar field in the 
chiral multiplet is
\begin{equation}
I_{\phi} =\frac{(t/y)^{\frac{q}{2 b}} (t y)^{\frac{q}{2 a}}}{\left(1-(t/y)^{\frac{1}{b}}\right) \left(1-(t y)^{\frac{1}{a}}\right)}
\end{equation}

Note that the two single particle indices that we have found only depend on the two parameters
$(ty)^{1/a}$ and $(t/y)^{1/b}$. Thus, the following redefinition of the fugacities
\begin{equation} \label{eq:redef}
t \to t^{\frac{a+b}{2}} y^{\frac{a-b}{2}}\quad, \quad 
y \to t^{\frac{a-b}{2}} y^{\frac{a+b}{2}}
\end{equation}
gives ${\cal I}_{\rm sphere} = {\cal I}_{\rm squash}$. The transformation \eqref{eq:redef} does not modify the physical content of the index,
because the fugacities are, a priori, arbitrary parameters. The only constraints come from the requirement of convergence
of the index, and are given by $ty<1$ and $t/y>1$ \cite{Kinney:2005ej}. Of course, the latter are preserved by
equation \eqref{eq:redef} for positive $a$ and $b$.

On the squashed sphere, the gaugino equation \eqref{gaugionosolns} gives
\begin{eqnarray}
\lambda_1(\theta)& =& \frac{e^{\frac{i \theta }{2}} }{\sqrt 2}h(\theta) \left(\frac{ m}{b \sin \theta}+i\frac{ n}{a\cos \theta}\right)
\, , \quad
\lambda_2(\theta) =-\frac{e^{-\frac{i \theta }{2}} }{\sqrt 2}h(\theta) \left(i \frac{ m}{b \sin \theta}+\frac{ n}{a\cos \theta}\right) \nonumber \\
E &=& -\frac{n}{a}-\frac{m}{b} 
\quad\quad\quad\quad\quad\quad\quad\quad \, \, \, ,\quad\quad
h'(\theta) = f(\theta) h(\theta) \left(\frac{n}{a}\tan\theta-\frac{m}{b}\cot \theta \right)
 \nonumber 
\end{eqnarray}
The solution for $\lambda$ is square integrable if $m,n \leq 0$, but we exclude the mode $(m,n)=(0,0)$ because it is
identically vanishing. The sum over the gaugino states gives
  \begin{equation}
   I_\lambda=-\left(
\sum_{m=0}^{-\infty} \sum_{n=0}^{-\infty} t^{-\frac{n}{a}-\frac{m}{b} }y^{\frac{n}{a}-\frac{m}{b}} - 1\right)
=-\frac{(t/y)^{\frac{1}{b}} (t y)^{\frac{1}{a}}-(t/y)^{\frac{1}{b}}-(t y)^{\frac{1}{a}}}{\left(1-(t/y)^{\frac{1}{b}}\right) \left(1-(t y)^{\frac{1}{a}}\right)}
\end{equation}

For the gauge bosons the equations (\ref{eq:BPSgauge}) become
\begin{eqnarray}
\begin{split}
\frac{\csc  \theta  \sec  \theta \mathcal{B}_{\theta }}{a b}+\frac{\mathcal{E}_{\theta }}{f}&\equiv y(\theta )\\
\frac{\csc  \theta  \mathcal{B}_{\alpha }}{b f}+\frac{\sec  \theta  \mathcal{E}_{\alpha }}{a}&= -i \sin  \theta \, y(\theta )\\
\frac{\sec  \theta  \mathcal{B}_{\beta }}{a f}+\frac{\csc  \theta  \mathcal{E}_{\beta }}{b}&= i \cos  \theta \, y(\theta )
\end{split}
\end{eqnarray}
  After applying the equations of motions 
 \begin{eqnarray}
\begin{split}
-i f^2\left(\frac{m \sec ^2\theta \mathcal{E}_{\alpha }}{a^2}+\frac{n \csc ^2\theta  \mathcal{E}_{\beta }}{b^2}\right)+\frac{\mathcal{E}_{\theta } f'}{f}-2 \cot 2 \theta  \mathcal{E}_{\theta }-\mathcal{E}_{\theta }' &=  0
 \\
\frac{i n \csc ^2\theta  \mathcal{B}_{\alpha }}{b^2}-\frac{i m \sec^2\theta  \mathcal{B}_{\beta }}{a^2}+E \mathcal{E}_{\theta }&= 0
 \\
-\frac{i n \csc ^2\theta  f^2 \mathcal{B}_{\theta }}{b^2}+\tan  \theta  \mathcal{B}_{\beta }+E f^2 \mathcal{E}_{\alpha }-\frac{\mathcal{B}_{\beta } f'}{f}+\mathcal{B}_{\beta }'&= 0
 \\
-\frac{i m f^2 \sec ^2\theta  \mathcal{B}_{\theta }}{a^2}+\cot  \theta  \mathcal{B}_{\alpha }+E f^2 \mathcal{E}_{\beta }+\frac{\mathcal{B}_{\alpha } f'}{f}-\mathcal{B}_{\alpha }'&= 0
\end{split}
\end{eqnarray}
   and the Bianchi identities
 \begin{eqnarray}
\begin{split}
i m \mathcal{B}_{\alpha }+i n \mathcal{B}_{\beta }+\mathcal{B}_{\theta }'&= 0
\\ 
E \mathcal{B}_{\beta }+i m \mathcal{E}_{\theta }-\mathcal{E}_{\alpha }'&= 0
 \\
E \mathcal{B}_{\alpha }+i n \mathcal{E}_{\theta }-\mathcal{E}_{\beta }'&= 0
 \\
E \mathcal{B}_{\theta }+i n \mathcal{E}_{\alpha }-i m \mathcal{E}_{\beta }&= 0
\end{split}
\end{eqnarray}
 we find
\begin{equation}
\left(\frac{\partial_{\theta}}{f(\theta)} -\left(\frac{m}{b}-\frac{1}{f(\theta}\right) \cot \theta +\left(\frac{n}{a}-\frac{1}{f(\theta)}\right) \tan\theta \right)
y(\theta) = 0 
\quad,
\quad
E =- \frac{n}{a} -\frac{m}{b} 
\end{equation}
  and the index is
  \begin{equation}
   I_{B}^{(V)}=
  \sum_{m=1}^{\infty}  \sum_{n=1}^{\infty} t^{ \frac{n}{a} + \frac{m}{b}} y^{  \frac{n}{a}-\frac{m}{b} } 
  =\frac{(t/y)^{\frac{1}{b}} (t y)^{\frac{1}{a}}}{\left(1-(t/y)^{\frac{1}{b}}\right) \left(1-(t y)^{\frac{1}{a}}\right)}
  \end{equation}

\section{Example: reducing $4d$ indices to $3d$ partition functions} \label{sec:reduction}

In this section we revisit the reduction of the four dimensional superconformal index 
to the three dimensional partition function
\cite{Kapustin:2009kz,Jafferis:2010un,Hama:2010av}. We will show that
 the reduction follows very easily, and the same argument can be generalized to
other dimensions.
The example of the round sphere can be found in
\cite{Dolan:2011rp,Gadde:2011ia,Imamura:2011uw}.

For concreteness, we consider the index on $S_b^3 \times S^1$ and show that it reduces to the
three-dimensional partition function ${\cal Z}_{S^3_b}$  by dimensional
reduction.
In four dimensions we consider the multi-particle index for a chiral and a vector multiplet,
that takes into account all the multi-trace gauge invariant combinations. 
The multi-particle index can be found by taking the Plethystic exponential
of the single particle index \eqref{eq:Isp}
\begin{equation}
{\cal I}_{m.p.} = \text{Exp}
\left[
\sum_{k=1}^{\infty} \frac{{\cal I}_{s.p.}(t^k,y^k,f^k,g^k)}{k}
\right]
\end{equation}
Comparing to equation \eqref{eq:Isp}, we have added two more parameters to the single particle
index: the fugacity $f$ for the internal flavor symmetries and the one $g$ for the gauge symmetry.
In the rest of this section we consider only the $f,y \to 1$ limit.

We start by looking at the contribution of the chiral multiplet.
As we already pointed out the fields contributing to the index on $S_{b}^3\times S^1$ are  the particle
$\phi$ for the bosonic component and the antiparticle  $\psi^{\dagger}$ for the fermionic component.
If one component is in the  $\rho$ representation of the gauge group, than the other component is in the $\bar \rho$.
By recalling the single particle result
\be
{\cal I}_\Phi = {\cal I}_\phi + {\cal I}_\psi = \sum_{BPS} \left( t^{E_\phi} g^\rho - t^{E_\psi} g^{-\rho} \right)
\ee
the multi-trace contribution to the superconformal index from a chiral multiplet in the $\rho$ representation of the gauge group is
\begin{equation}
\text{Exp}
\left[\sum_{k=1}^{\infty} \frac{1}{k} \sum_{BPS} \left( t^{k E_{\phi}+i k \sigma \rho}-t^{k E_{\psi}-i k \sigma \rho} \right)
\right]
\end{equation}
that becomes
\begin{align} \label{impo}
\prod_{BPS} \frac{1-t^{E_{\phi}+i  \sigma \rho }}{1-t^{E_{\psi}-i  \sigma \rho }} 
\,\, \stackrel{t\to 1}{\longrightarrow} \,\,
\prod_{BPS} \frac{E_{\phi} +i  \sigma \rho}{E_{\psi} -i  \sigma \rho}
\end{align}
where we identified the chemical potential for the gauge group $g$ with $t^{i\sigma}$, where $\sigma$ is the
 solution  to the three-dimensional saddle point equations (or to the four-dimensional zero energy
supersymmetry equations), which set $\sigma$ to a
 constant \cite{Kapustin:2009kz}.\footnote{The reason for setting $g=t^{i\sigma}$,
in our language, is the following.
Till now, we solved the BPS equations in a vanishing gauge background, because we know that the gauge
representation can be associated to another chemical potential in the index (also see footnote \ref{fn:gauge}).
However, we could have solved the BPS
equations in the $\sigma\neq 0$ background and obtain that the energies are $E^\prime = E+i \sigma\rho$. A comparison
of the two methods shows that $g$ goes as $t^{i \sigma}$ when the time circle shrinks.}

The product in \eqref{impo} ranges over the set of BPS states. As we have seen, this set is labeled by
the Cartan subgroup of the three-dimensional
  isometry group,  which  in the case at hand consists of the two $U(1)$ symmetries $U(1)_\alpha$
and $U(1)_\beta$ that rotate the Hopf angles independently.
If we identify the fugacity $t$ with $e^{-\tau}$, where $\tau$ is the period of the time direction, then the
limit in \eqref{impo} corresponds to shrinking the time circle to zero size, i.e. to dimensional reduction.
Indeed the right hand side of (\ref{impo}) is the one loop exact contribution of the
chiral multiplet to the three-dimensional partition function found in 
\cite{Hama:2011ea}.
The energies $E_\phi$ and $E_\psi$ of the BPS states in four dimensions,
obtained with the procedure explained in section \ref{sec:4d}, become the
eigenvalues of the unpaired states in the three dimensional case. 
An analogous derivation can be performed for the vector multiplet.

We expect that our correspondence and the reduction are more general
than shown here and that they apply generically to $\mathcal{M}_{d-1}
\times S^1 \rightarrow \mathcal{M}_{d-1}$ \footnote{ It would be
  interesting to study the same correspondence between the states on
  $\mathcal{M}_{d-n} \times T^n$ and the ones on $\mathcal{M}_{d-n}
  \times T^{n-1}$.}, provided at least two real supersymmetries are
preserved.  The result \eqref{impo} should apply to any
$(d-1)$-dimensional theory, if its field content may be derived by
dimensional reduction of a corresponding $d$-dimensional model. The
$(d-1)$-dimensional saddle points and the quantum corrections may be
derived by the $d$-dimensional analysis, but in the full partition
function there may be an additional contribution, denoted $S_\ast$ in
\eqref{eq:Ztot}, due to a classical term which does not have an uplift
to $d$-dimensions. This is the case, for instance, for the
Chern-Simons term in the three dimensional case.  However, once the
$(d-1)$-dimensional action is known, one can plug the saddle point
configuration in it and obtain also the classical term.

It is interesting to compare with the known results in the literature.
To the best of our knowledge, this is the first time that the
superconformal index on a squashed sphere is computed explicitly. Of
course, because it is identical to the one on the round sphere up to a
redefinition of the fugacities, one can consider reducing the latter
to the partition function on the squashed sphere. This is usually done
by taking an ad hoc limit instead of the one in \eqref{impo}
\cite{Dolan:2011rp}.  Namely, we can reinterpret those results by
stating that one can squash the chemical potentials without affecting
the physical meaning of the index, and then take the natural limit
$t\to 1$ to shrink the time circle. The necessary redefinitions are
not known in general, and we believe that our results offer a very
clean physical interpretation and can be easily generalized.

\section{The conjecture in other dimensions and manifolds} \label{sec:extensions}

From the discussion in section \ref{sec:reduction} we see that our results can be more general
than stated until now. We propose that the same one-to-one map described there
holds in more general cases, like in other dimensions, manifolds and for extended
supersymmetric theories.

Localization on a three-sphere does not give rise to any non-perturbative (instanton or monopole)
 contribution, and this
is in full agreement with the BPS correspondence we have proposed. However the localizing
term in different dimensions can lead to a sum over the instantons as happens, for instance,
on the four-sphere. If our argument can be applied also in that case, the five-dimensional
BPS equations should contain all the quantum information also about the non-perturbative states.

In section \ref{sec:4d} we have mostly focused on a
three-dimensional manifold whose Cartan subgroup is $U(1)^2$, and thus there are two
well-defined quantum numbers, one can break the Cartan to $U(1)$ and still preserve two real
supersymmetries. In this case one has
only one integer quantum number to sum over, and the BPS conditions will give constraints on
its range.

\section{General partition functions via an uplift to an index} \label{sec:uplift}

We have observed above that the reduction of the superconformal index
on $\mathcal{M}_{d-1} \times S^1$ to the partition function on
$\mathcal{M}_{d-1}$ highlights the relation between the BPS states in
$d$ dimensions and the $d-1$ dimensional unpaired states.
Equivalently one can obtain the three dimensional partition function
on $\mathcal{M}_{d-1}$ by uplifting the supersymmetry from
$\mathcal{M}_{d-1}$ to $\mathcal{M}_{d-1} \times S^1$.  The
$d$-dimensional Killing spinors are independent from the $S^1$ and the
$d$ dimensional unpaired states are preserved by shrinking the circle.
Even if this procedure is similar to the reduction explained in
section \ref{sec:4d} it is interesting to investigate the problem in
this way because it shows the relation of our construction and
localization.  Indeed the $d-1$-dimensional saddle point equations
coincide with the zero energy equations of the $d$-dimensional
problem.  We now exploit this fact to simplify the computation of the
exact partition function itself.

Consider a $d$-dimensional field theory ${\cal F}_d$ and its
dimensional reduction to ${\cal F}_{d-1}$, which preserves the same
amount of supersymmetry.\footnote{Actually, the action for ${\cal
    F}_{d-1}$ may contain terms without an uplift to $d$
  dimensions. As we already stressed our results also hold in those
  cases.}  We can place ${\cal F}_{d-1}$ on a curved manifold ${\cal
  M}_{d-1}$ and localize the corresponding path integral to an at most
finite dimensional integral by picking two real conserved supercharges
and solving the corresponding equation $|\delta \psi_{d-1}|^2=0$,
where $\psi$ is any fermion of the theory. This is the same as picking
the uplifted supercharges on ${\cal M}_d$ and solving for
\be \label{eq:saddles} |\delta\Psi_d|^2\Big|_{E=0}=0 \ee where $\Psi_d
\sim \Psi_{d-1} e^{E t}$ is the set of fermions in the ${\cal F}_d$
theory, and gives the loci that solve the saddle point equations in
the path integral. Denote the latter by $\Phi_\ast$ and the classical
action $S(\Phi_\ast)\equiv S_\ast$. The exact path integral on ${\cal
  M}_{d-1}$ is now given by \be \label{eq:almost} Z_{{\cal M}_{d-1}}
\sim \int \left[d\Phi\right] \,\, e^{-S_\ast} \frac{{\rm Pf}
  D_F}{\sqrt{\det{D_B}}} \ee where $[d\Phi]$ is the measure over the
loci $\Phi_\ast$, and in general $D_F$ and $D_B$ are respectively a
first order and second order differential operator derived by a
$(d-1)$-dimensional Q-exact action. Notice that we did not compute any
Q-exact action, so we do not know the explicit form of $D_F$ and
$D_B$, but we know that $\Phi_\ast$ are their zero modes.  In general,
we should find the spectrum of their eigenvalues around the solutions
of \eqref{eq:saddles}, and it turns out that many of them simplify
between the numerator and the denominator in \eqref{eq:almost} due to
supersymmetry.  The ones that do not simplify are obtained with the
procedure explained in section \ref{sec:4d}.

To summarize we can derive the spectrum of eigenvalues necessary to
compute the exact partition function in $d-1$ dimensions
\eqref{eq:almost} by finding the energy eigenvalues from a
corresponding set of first order differential operators in $d$
dimensions.  We do not need the Lagrangian giving the equations of
motion for ${\cal F}_d$, but only the supersymmetry transformations of
the matter multiplets that appear there.  This means that we only need
the uplift of the conserved supercharges, without worrying about the
uplift of the Lagrangian.

\section*{Acknowledgments}
We thank O.~Aharony, M.~Berkooz, L.~Girardello, K.~Intriligator, Z.~Komargodski,
J.~McGreevy, 
M.~Petrini, J.~Song,  D.~C.~Stone, D.~Thompson, A.~Tomasiello
and A.~Zaffaroni for useful discussions
and comments on the manuscript.
A.M. acknowledges funding by
the Durham International Junior Research Fellowship.
M.S. is a Feinberg postdoctoral fellow at the Weizmann Institute of Science.
M.S. would like to thank the University of Milano-Bicocca and UCSD for their kind hospitality
during the completion of this paper.

\appendix
\section{More on the gauge field contribution}

In this Appendix we show a different approach to compute the contribution of the gauge
field to the index. We focus on the four-dimensional case of $S^1 \times S^3_b$, and the
round sphere can be obtained by taking the $a,b \to 1$ limit.

As explained in the main text, our method relies on finding a map from
the bosonic modes to the fermionic ones such that their contributions
to the index cancel out, and the unpaired modes are identified with
the BPS states. We have seen that the normal modes of the gauge field
strength can be mapped to the gaugino modes. Among the unpaired modes
of the gauge field strength, those which also satisfy the Maxwell
equations are the BPS modes. Here we offer another interpretation for
the latter.

Besides the map between the gauge and the gaugino,
we can find another map that relates a mode of the gauge field strength to a fermion
with the same $(\Xi,H,\tilde J_3)$ quantum numbers
\bea
\delta\chi = \tilde \zeta \tilde\sigma^\mu v_\mu
\eea
The field $\chi$ is a pure supergauge field, i.e. it is set to zero in the Wess-Zumino gauge.
For this reason it does not belong to the Hilbert space and every gauge field such that
\bea \label{eq:gaugefix}
\tilde \zeta \tilde\sigma^\mu v_\mu = 0
\eea
can contribute to the index. The solution to this equation is
\be
\begin{split}
v_1 &= 0 \\
v_2 &= Y(\theta) e^{E t+i (n \alpha +m \beta )} \\
v_3 &= i a \frac{\sin{\theta} \cos{\theta}}{f(\theta)} Y(\theta)  e^{E t+i (n \alpha +m \beta )} \\
v_4 &= -i b \frac{\sin{\theta} \cos{\theta}}{f(\theta)} Y(\theta) e^{E t+i (n \alpha +m \beta )}
\end{split}
\ee
where the first line is a gauge choice and $Y(\theta)$ is to be determined.
Then the two equations $\sigma^{\mu\nu} F_{\mu\nu} \zeta=0$
for $Y(\theta)$ give
\be
E = -\frac{n}{a}-\frac{m}{b}
\ee
for $n,m \leq 1$. This is the same result that we obtained in section \ref{sec:example1}.
Notice that equation \eqref{eq:gaugefix} is satisfied by the pure gauge configuration
$v_\mu=\partial_\mu \Phi$ that appears in equation \eqref{solgaugino}.

\bibliographystyle{JHEP}
\bibliography{bpsref}

\end{document}